# A Multi-Layer SEU Mitigation Strategy to Improve FPGA Design Robustness for the ATLAS Muon Spectrometer Upgrade

Xueye Hu, Jinhong Wang, Reid Pinkham, Suen Hou, Thomas Schwarz, and Bing Zhou

*Abstract*— We present a multi-layer single-event upset mitigation strategy implemented in a low-cost Xilinx Artix-7 FPGA. The implementation is targeted for a trigger data router for the ATLAS muon spectrometer upgrade. The mitigation strategy employs three layers of protection to improve overall FPGA design robustness: use of triple-modular redundancy for FPGA fabric logic and embedded soft-error mitigation in the first layer; further enhancement with multi-boot FPGA reconfiguration across multiple copies of configuration memory in the second layer; and FPGA power cycling and configuration memory re-initialization in the third layer. The effectiveness of this scheme has been evaluated at two different neutron facilities, LANSCE and NCSR "Demokritos", with 800 MeV and 25 MeV beam energies, respectively. Testing was performed with a similar configuration to that planned for final operation. We discuss the testing strategy and summarize the test results to estimate the expected data loss over 10 years of operation in the ATLAS experiment.

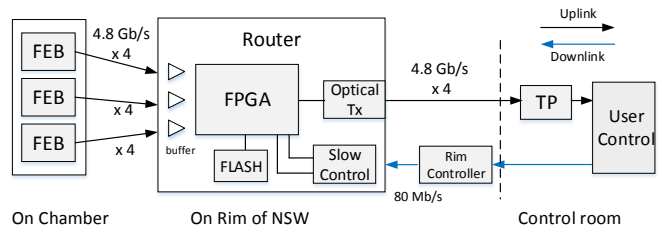

Fig. 1. Illustration of signal flow of the primary trigger path in the ATLAS NSW detector and the role of Router in handling the serial streams.

## I. INTRODUCTION

THE muon spectrometer forms the outer part of the ATLAS detector and is designed to detect charged particles exiting the barrel and end-cap calorimeters and to measure their momentum in the pseudo-rapidity range $|\eta| <$ 2.7. It is also designed to trigger on these particles in the region $|\eta| <$ 2.4 [1]. To benefit from the expected high luminosity provided by the Phase-I upgrade of the LHC, the first station of the ATLAS muon end-cap system (Small Wheel, SW) will be replaced with a "New Small Wheel" (NSW) detector which can better handle the increased data rates and radiation [2]. For example, the expected raw data rate from ATLAS at the current luminosity is up to 1 petabytes per second, and will increase as luminosity increases. ATLAS utilizes an advanced trigger system to reduce the data flow to a manageable level, and the trigger path in the NSW will play a critical role in retaining excellent muon triggering selectivity under the large background at the high luminosity LHC.

This paper discusses the design of a trigger data packet router which is used in the data acquisition system of the Small Thin Gap Chambers (sTGC) of the NSW detector. A simplified block diagram of the primary trigger path in the NSW is shown in Fig.1. In general, it is made up of high-speed uplinks for immediate trigger data transmission and relatively low-speed downlinks for slow control and monitoring. The trigger packet router (Router) in the uplinks serves as a fast switch-yard between the front-end boards (FEB) and the back-end trigger processor (TP). Each router handles a maximum of twelve 4.8 Gb/s incoming links from three FEBs, suppresses NULL packets and passes over data packets to the TP via four 4.8 Gb/s optical links [3-4].

The high-speed links through the Router handle the trigger data flow from several sTGC chambers so the router must operate with high reliability to prevent data loss even in the presence of radiation effects at the location of the NSW detector. In the implementation of the Router, a commercial SRAM-based field programmable gate array (FPGA) from the Xilinx low-cost Artix-7 family (XC7A200TFFG1156) was chosen in favor of a custom ASIC. The use of a commercial FPGA offers reduced cost, a short development cycle, and design flexibility. However, FPGAs are vulnerable to radiation-induced effects and care must be taken to mitigate these effects.

An explicit review of both the radiation-induced effects in modern FPGAs and associated mitigation strategies is discussed in [5-9]. There are also some studies specifically on the Xilinx 7 series FPGAs [10-13] which cover mostly the moderate or high performance families: e.g. Kintex or

Manuscript received on Jan. 18, 2019, this works is supported by the Department of Energy under contracts DESC0007857 and DE-AC02-98CH10886.

S. Hou is with Academia Sinica, Taipei; J. Wang, R. Pinkham, X. Hu, T. Schwarz, B. Zhou are with Department of Physics, University of Michigan, Ann Arbor, MI, 48109, US. (email: jinhong@umich.edu; xueyehu@umich.edu).

Ultrascale. Relevant applications in the high-energy physics experiments can be found in [13-16].

In this paper, we present a multi-layer SEU mitigation strategy to improve the robustness of the Router in the harsh environment of the ATLAS NSW detector. This work is structured as follows. Section II outlines the radiation environment that the Router will be exposed to in the ATLAS experiment for the next decade and the primary radiation effects expected. Section III covers the implementation of the mitigation techniques. Section IV presents a test system to evaluate Router performance under radiation exposure. Experimental results from two different neutron beam facilities are shown in Section V. Discussions are given in Section VI and a conclusion is made in Section VII.

## II. EXPECTED RADIATION ENVIRONMENT AND EFFECTS

### A. Radiation Levels at the rim of ATLAS NSW detector

The Router will be located on the outer rim of the NSW detector in the ATLAS muon spectrometer. At this location there is radiation originating directly from the proton-proton interactions as well as secondary radiation from activated materials and nuclear interactions throughout the detector. The radiation includes high energy photons, neutrons, and charged particles. Due to this radiation all commercial off-the-shelf (COTS) components on the Router, including the Artix-7 FPGA, must be evaluated for radiation tolerance to ensure compliance with the ATLAS Radiation Test Criteria (RTC) [17]. The RTC document states that radiation tolerance criteria can be calculated from the product of the simulated radiation level (SRL) listed in Table I and the corresponding safety factors (SF) listed in Table II [18]:

$$RTC = SRL \times SF_{sim} \times SF_{ldr} \times SF_{batch} \quad (1)$$

The safety factors account for simulation uncertainties ($SF_{sim}$), low-dose-rate effects ($SF_{ldr}$), and lot-to-lot variation ($SF_{batch}$). The numbers in Tab. I are calculated for the NSW after 10 years at an LHC luminosity of $5 \times 10^{34}$ cm$^{-2}$s$^{-1}$.

According to equation (1), the Artix-7 FPGA on the Router must tolerate: 16 Gy $\times$ 1.5 $\times$ 5.0 $\times$ 4.0 = 480 Gy (48 kRad) of total ionizing dose (TID), and a total fluence of: $1.3 \times 10^{11} \times 2.0 \times 1.0 \times 4.0 = 1.04 \times 10^{12}$ p/cm$^2$ for single event effects (SEE), over the course of 10 years operation.

### B. Radiation Effects on the Router FPGA

There are multiple ways that radiation can affect the operation of electronics. The damage is normally divided into three different categories: TID, displacement damage (Non-ionizing Energy Loss: NIEL), and SEE. NIEL effects are not considered here as modern CMOS integrated circuits are insensitive to displacement damage [17].

*1) Total Ionizing Dose*

Semiconductors are susceptible to failures due to the accumulation of charge over time. Incoming high energy photons can cause electron-hole pair generation which produces excess free charge if the pairs do not recombine. The excess charge can lead to errors in the internal logic, or cause changes in threshold voltage or leakage current. The amount of radiation dose that a device can tolerate before failing to meet published parameter specifications is called TID [5]. The Artix-7 FPGA was qualified it up to 550 kRad with Co-60 γ source at BNL's facility [19], which is far beyond the required 48 kRad.

TABLE I
NSW SIMULATED RADIATION LOADS (SRL)

| Type | Inner Rim (R=1m) | Outer Rim (R=5m) |
|---|---|---|
| TID (γ) | 780 Gy | 16 Gy |
| NIEL (fast neutrons) | $2.3 \times 10^{13}$ n/cm$^2$ | $7.3 \times 10^{11}$ n/cm$^2$ |
| SEE (protons) | $4.2 \times 10^{12}$ p/cm$^2$ | $1.3 \times 10^{11}$ p/cm$^2$ |

TABLE II
RADIATION SAFETY FACTORS FOR NSW COTS ELECTRONICS

| Safety Factor | Type | Value | Notes |
|---|---|---|---|
| $SF_{sim}$ | TID | 1.5 | |
| | NIEL | 2.0 | |
| | SEE | 2.0 | |
| $SF_{ldr}$ | TID | 5.0 | COTs, no control for low-does-rate effects |
| | NIEL | 1.0 | |
| | SEE | 1.0 | |
| $SF_{batch}$ | ALL | 4.0 | Unknown COTs batches |

*2) Single Event Effects*

In addition to cumulative effects (TID, NIEL), the interaction of high-energy particles with electronic elements in integrated circuits can cause immediate effects that are collectively called "single event effects" (SEE). When a high-energy particle passes through the silicon substrate of a device, charged particles are created as the result of sub-atomic particle collisions. These particles are generated by an ionization trail along the path of the incoming particle. The major circuit effects from high-energy particles include transient current pulses, change in memory values (bit flips or SEUs), and latch-ups. These effects can be divided into two broad categories: recoverable errors (single-event transients, single-event upsets, single-event function interrupts) and non-recoverable errors (single-event latch-up, single-event burnout, single-event gate rupture). For an FPGA, these effects can occur in fabric logic elements, configuration memory, and internal proprietary control elements. They may disrupt proper functionality of FPGA design and lead to unexpected behavior of system communication.

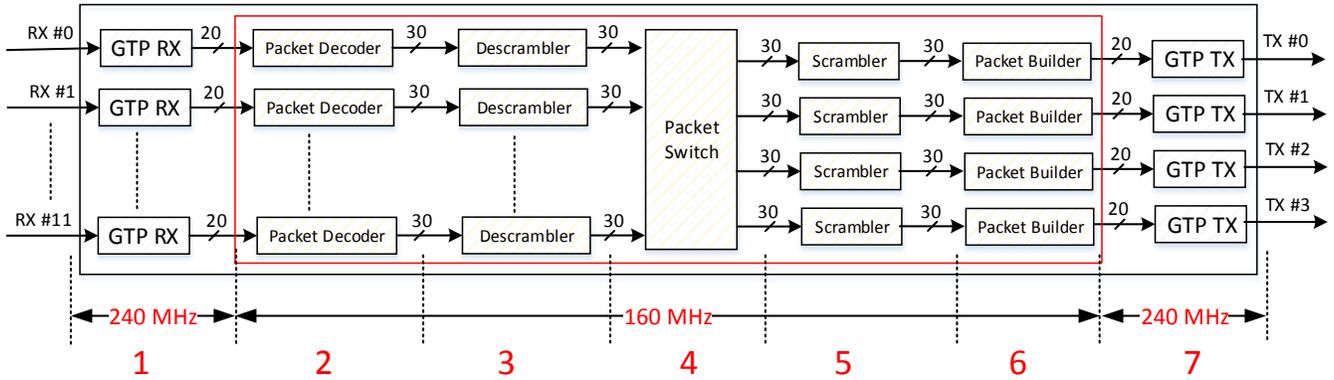

Fig. 2. Processing of serial streams inside the FPGA of the router: stage 1 and 7 correspond to the GTP transceivers; stages 2-6 represent the user logic.

## III. MULTI-LAYERS SEU MITIGATION ARCHITECTURE

An illustration of the data flow inside the Router is shown in Fig.2 which shows seven stages of the serial signal flow. In Stage 1, the twelve lines of 4.8 Gb/s serial stream from FEBs are received by the Xilinx GTP receiver (GTP RX). The RX groups the data in 20-bit packets which are buffered in stage 2 and descrambled in stage 3 to reconstruct the original 30-bit packets from the FEB. A FEB packet is 30-bits, consisting of a 4-bit header followed by a 26-bit payload. There are two header patterns: "1010" or "1100" for data and a NULL distinction respectively. Headers are used for quick data switching and NULL suppression in the Router. The 26-bit payload is scrambled on the FEB to keep DC. In Stage 4, NULL packets are dropped and data packets are forwarded. Data packets are scrambled again in Stage 5 and assembled in TP packet format in Stage 6. Finally, the packets are shifted out via four Xilinx GTP transmitters (GTP TX) in Stage 7. In the whole signal flow from Stage 1-7: Stage 1 and 7 include both hard IP cores (e.g. the GTP transceivers, clock managers) and FPGA fabric logic for control and initialization; Stage 2-6 are made up of pure fabric logic elements.

### A. First Layer: TMR and SEM Controller

Operations from Stage 1-7 can be corrupted by SEUs from different parts of the FPGA, e.g. the hard IP cores, fabric logic elements, or configuration memory. In the first protection layer, we apply direct SEU mitigation to fabric logic elements and configuration memory. The logic elements are directly accessible to designers while Xilinx offers SEU mitigation tools for the configuration memory. SEUs in these two parts also account for most of the SEU corruptions in an FPGA design. Other parts of the FPGA, such as the hard IP cores, are either invisible to designers or have no associated resources available for direct SEU mitigation. SEUs in these parts are handled indirectly in a higher layer.

For critical logic sections that handle control and initialization, their design states are protected by triple modular redundancy (TMR) [20]. We implemented TMR using three voters for the strongest protection. The goal is to retain a steady data path from Stage 1-7, while allowing bit upsets in the data stream to be recovered using the redundancy of data from multiple detector layers.

Configuration memory are storage elements used to configure the function of the design loaded into the device. They are physically distributed across the entire device and represent the largest number of memory bits in the FPGA. Fortunately, configuration memory bits are already protected with Error-Correction Code (ECC) and Cyclic Redundancy Check (CRC) and only a fraction of these memory bits are essential to proper operation. However, there is still the possibility that SEUs happen in these essential bits and thus change the behavior of the design. Accumulated upsets will eventually result in functional failures. In the first layer, upsets are addressed by a Soft Error Mitigation (SEM) tool from Xilinx [21] which monitors the integrity of the configuration memory and can fix up to two bits upset simultaneously.

### B. Second Layer: Multi-boot Auto Reconfiguration

Even with TMR and SEM controls, there is still some chance for more than 2-bits to have an upset – a Multi-bit upset (MBU), which disrupts the protection in the first layer. An MBU can be recovered by reconfiguration via the slow control path in Fig.1. However recovery using slow control may take several minutes during which the circuit may be in an unexpected state or produce incorrect results. For the Router operation, this means that there will be some time during which the TP will lose trigger information from a whole detector sector. To minimize the total time of data loss in this case, we implement a second mitigation layer: an on-board Multi-boot Auto Reconfiguration (mBAR), which utilizes an on-board flash memory to hold the FPGA configuration file thus avoiding the need of transmission through the slow control path.

We adapt the same idea as the MultiBoot scheme proposed by Xilinx except that there is no golden image in the Router on-board flash memory [22]. The flash is based on Cypress® 65 nm MirrorBit® Technology with Eclipse™ Architecture and it has automatic ECC for single bit error correction. Additional TID tests certified that this flash can tolerate the accumulated radiation on the Router board [19].

The mBAR is performed by mapping multiple copies of

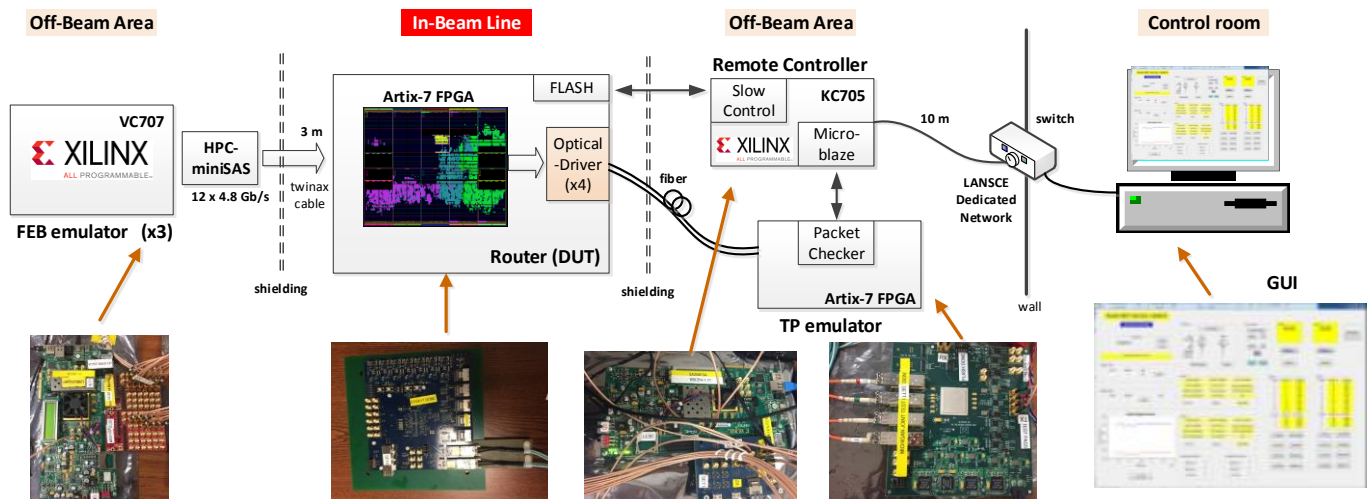

Fig. 5. Block diagram of Router SEE evaluation system. See text for explanation.

Router firmware design into the 256Mb Cypress flash. We utilized the bitstream compression option from the Xilinx FPGA to reduce the size of the configuration file. The option works by writing identical configuration frames once instead of writing each frame individually. As a result, the size of compressed Router firmware is reduced from about 78 Mb to around 32 Mb, permitting 6 copies to be loaded in the flash memory. A complete firmware reload from the flash memory can be done in several seconds, compared to several minutes using the slow control system. The stored copies in the flash are independent of each other and are queued in sequence to be loaded into the FPGA in the second layer. Firmware reload via the slow control path will be considered only when all copies in the flash are corrupted.

### C. Third Layer: Router Power Cycling

Most the upsets can be addressed in the first two protection layers. However, upsets in the SEM controller itself, or upsets in internal circuits of the hard IP cores may put the IPs in an unknown state that cannot be fixed by the protection schemes in the first two layers. These abnormal states can be detected by monitoring logic in the FPGA and the integrity of Router packets from the TP in Fig.1. The backend controller collects any abnormal states of the Router using the slow control interface and feedback from the associated TP. In the case of constant failures, an mBAR will be issued first. If the errors persist, a power cycle will be performed. The currents, voltages, and temperatures on the Router are monitored via the slow control interface. Power cycling will also be applied in case of an alarm due to abnormal currents on the Router. In summary, this is the last protection layer, and any failures cannot be corrected in the prior two layers will be addressed here, including non-recoverable SEEs.

## IV. ROUTER SEU TEST SYSTEM

### A. Hardware Implementation

We evaluated the radiation performance of the Router with the test system as shown in Fig.5. The system emulates the data flow in the NSW detector and is divided in 4 sections, as follows:

1) *"FEB emulator"*

A Router handles a maximum of 12 4.8 Gb/s serial inputs from three FEBs, and this input data is emulated by a Xilinx VC707 evaluation board. The serial outputs of the FEB packets are converted by a bridge board: converting a High Pin Count (HPC) connector from the VC707 to the miniSAS connector (HPC-miniSAS); and then transmitted through 3m twinax cable to the Router. The "FEB emulator" was placed in the off-beam area.

2) *"Router Prototype"*

The design under test (DUT) is a pre-production version of the Router, in which all functionalities are implemented. A Router decodes incoming serial streams, forwards selected packets to the "TP processor" via four optical fibers, and also communicates with the "Slow Controller" for control and monitoring. In addition, to accommodate the Aritx-7 FPGA radiation test, there is a 2-inch clear distance from the center of the FPGA to separate other active components. The radiation tests were performed by aligning the beam towards the center of the FPGA to ensure all radiation induced errors are sourced from the FPGA itself.

3) *"TP emulator"*

A custom board was built to emulate the TP processor, which checks the integrity of test frames from the Router. These packets originate from the FEB emulator and arrive at the TP through the uplinks. There are four optical channels on the "TP" emulator to decode the forwarded packets from the DUT accordingly. "TP emulator" is also in the off-beam area.

4) *"Remote Controller"*

A Xilinx KC705 FPGA evaluation board is deployed as the remote controller for the whole evaluation system. It collects the feedback status from the "DUT" and "TP emulator", and reports this monitoring information to a user in the control room via dedicated Ethernet link. In addition, the "remote controller" also responds to commands from the user and distributes control signals like reset, multi-boot trigger, power cycling, to associated units. All this is done with an embedded micro system (MicroBlaze) in the KC705 FPGA.

*B. Software Test Fixture*

A graphical user interface (GUI) was also created for automatic control and monitoring of the radiation characterization, as in Fig. 5. The GUI was developed via the MATLAB GUIDE tool [23]. It communicates with the KC705 controller board via TCP/IP server sockets to send commands and to read status information. This GUI also analyzes any Router SEU failure event in real time following the link loss criteria set in the software, as shown in Fig. 6. The criteria have three sections, reflecting the status of the DUT from two aspects: serial link status (in DUT) and the integrity of the payloads (in both DUT and TP emulator).

Serial link status is checked from the reported header positions of the Router, which represents the relative position of FEB packets in a link from the recovered raw packets in Stage 1-2, in Fig.2. Inconsistent values of this criteria indicate the loss of synchronization in a link, possibly caused by SEUs.

The payloads are checked at the Router and the TP emulator, respectively. The former covers the data processing in Stages 1-3 as in Fig.2, as well as the checking logic inside Router; while the latter corresponds to that in Stages 4-7. The payload for a complete data packet is made up of 104 bits and are grouped into four 30-bit data frames. There are pre-assigned increments for every 8 bits of the 104-bit payload, and its integrity is reflected by the 13-bit flags, with bit "0" represents an error in the corresponding 8-bit group, bit "1" for none error observed. A unique identification number (ID) is also assigned for each Router and this information is passed over to the TP whenever there is no data forwarded to a TX in Stage 7. TP tracks the Router ID as part of the data integrity checks.

In the case that any of the three indicators discussed above show persistent errors longer than 5 minutes (a configurable value in the GUI), an SEU failure is reported and will add up the failure counter by 1.

The GUI panel also displays other real-time monitoring information for the evaluation system: e.g., current dissipation versus elapsed time, locking status of FPGA IP cores (clock manager, GTP transceiver), serial number of bit stream loaded, and accumulated failures. All this information is also logged for off-line analysis.

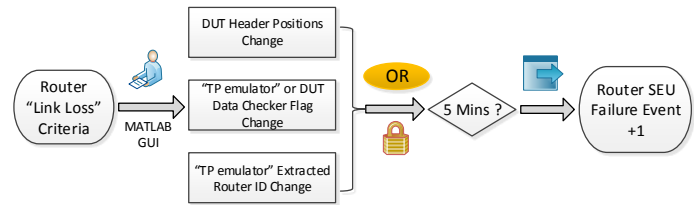

Fig. 6.  Flow diagram of link loss criteria set in Router SEU test system.

TABLE III
NEUTRON BEAM CHARACTERISTICS IN NSCR AND LANSCE

|  | NSCR | LANSCE |
|---|---|---|
| Beam Energy | Up to 20 MeV | Up to 800 MeV |
| Beam Size | 4π | Collimated in 2 inch diameter |
| Flux (n/cm$^2$/s) | $3.41 \times 10^4$ - $1.36 \times 10^6$ | $1.39 \times 10^6$ |

## V. EXPERIMENTAL RESULTS

The Router radiation was performed at two different neutron facilities: one was at the Los Alamos Neutron Science Center (LANSCE), Los Alamos, USA, with the beam energy up to 800 MeV; the other was at the National Centre of Scientific Research (NSCR) "Demokritos", Greece, with a maximum energy of about 20 MeV. In each evaluation, two identical setups were run in parallel for cross verification. A comparison of the two beam characteristics is shown in Table III.

TABLE IV
TEST RESULTS FROM NSCR AND LANSCE

| Neutron Beam | | SET I | | | | SET II | | | |
|---|---|---|---|---|---|---|---|---|---|
| | | Flux (n/cm$^2$/s) | Fluence (n/cm$^2$) | Soft Failure | Hard Failure | Flux (n/cm$^2$/s) | Fluence (n/cm$^2$) | Soft Failure | Hard Failure |
| NSCR | RUN 1 | $3.41\times10^4$ | $1.46\times10^9$ | 0 | 0 | $1.46\times10^4$ | $6.25\times10^8$ | 0 | 0 |
| | RUN 2 | $3.92\times10^5$ | $1.55\times10^{10}$ | 3 | 0 | $2.02\times10^4$ | $4.21\times10^8$ | 0 | 0 |
| | RUN 3 | $4.98\times10^5$ | $1.80\times10^{10}$ | 4 | 0 | $5.37\times10^4$ | $1.94\times10^9$ | 0 | 0 |
| | RUN 4 | $1.36\times10^6$ | $2.83\times10^{10}$ | 9 | 0 | $5.74\times10^4$ | $2.27\times10^9$ | 1 | 0 |
| LANSCE | RUN 5 | $1.39\times10^6$ | $1.30\times10^{11}$ | 94 | 3 | $1.39\times10^6$ | $1.27\times10^{11}$ | 88 | 2 |
| | RUN 6* | $1.39\times10^6$ | $3.15\times10^{11}$ | 116 | 7 | $1.39\times10^6$ | $3.06\times10^{11}$ | 105 | 6 |

*Note: Checking criteria for RUN 6* at LANSCE was adjusted for final data loss estimation.

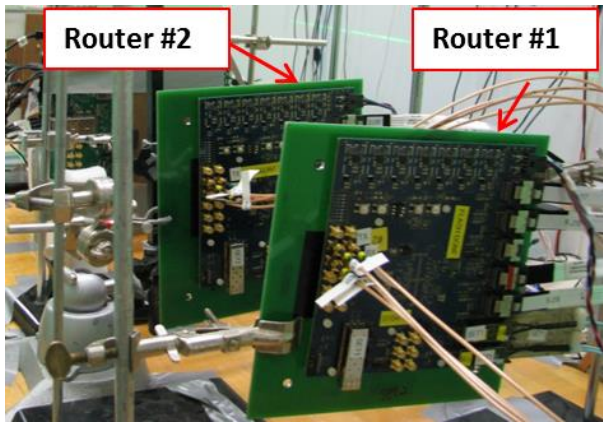

Fig. 7. Router SEE Radiation Test at LANSCE

In characterizing the SEU mitigation performance of the Router, we define two categories of failures: soft and hard. The rate of these failures in the Router were evaluated in terms of a cross section ($\sigma_{link}$). Soft failures are those recoverable with the mBAR scheme in the second protection layer, while hard failures are those requiring a power cycling in the third protection layer. The calculation of $\sigma_{link}$ is as $\sigma_{link} = n_{events}/ (\Phi_{fluence} \times \cos(\theta))$, in which $n_{events}$ is the count of the radiation-induced link breaks during the beam time, $\Phi_{fluence}$ is the fluence during testing beam time, $\theta$ is the angle of the test fixture relative to the beam (in following evaluation $\theta = 0$).

*A. NSCR "Demokritos" Test Results*

The neutron beam at NSCR has a $4\pi$ solid angle, thus the closer a DUT is to the beam outlet, the higher flux the DUT will experience. For the two sets, SET I and II, at the NSCR facility, SET I was placed directly at the beam outlet, which experienced the highest flux; while SET II was farther away to leave space for other test devices according to the beam test arrangement. In total, four test cycles were performed while increasing the outlet flux from $3.41\times10^4$ n/cm$^2$/s to $1.36\times10^6$ n/cm$^2$/s. Corresponding flux for SET II is from $1.46\times10^4$ n/cm$^2$/s to $5.74\times10^4$ n/cm$^2$/s. A total of 16 soft errors and zero hard error were observed in SET I, whereas only one soft error was observed in SET II. Test results are summarized in Table IV. The soft error cross section $\sigma_{link}$ for SET I and II was estimated to be $2.53\times10^{-11}$ cm$^2$/n and $1.9\times10^{-10}$ cm$^2$/n, respectively. No hard failures were observed at NSCR. In addition, the flux of RUN 4 is comparable with that at LANSCE, and the corresponding soft failure cross section was around $3.18\times10^{-10}$ cm$^2$/n.

*B. LANSCE Facility Test Results*

For the tests at LANCE, the flux of the neutron beam was constant at $1.39\times10^6$ n/cm$^2$/s and uniform within the 2-inch collimated area. Two parallel setups were used and placed close together to obtain a similar exposure as shown in the picture in Fig.7. Two runs were taken, denoted as RUN 5 and 6, whose data are shown Table IV. In RUN 5, identical checking criteria was used as that at NSCR, and calculations shows the soft error cross section is about $7.15\times10^{-10}$ cm$^2$/n and $6.93\times10^{-10}$ cm$^2$/n for SET I and II, respectively. Hard failures were first observed at LANSCE, and the cross section is on the order of $10^{-11}$ cm$^2$/n. Further analysis of monitoring logs from the GUI show that the three hard failures include 2 failures from the clock manager IP cores (e.g. PLL) and one from the SEM controller.

To estimate the final link loss in the detector operation, the failure checking criteria in Fig.7 were revised for RUN 6 and DUT data checking flags were removed. This is because only the frames passing from the 12 inputs to the 4 outputs is of interest in the experiment. As a result, the soft failure cross section for SET I and SET II in RUN 6 were estimated as $3.68\times10^{-10}$ cm$^2$/n and $3.43\times10^{-10}$ cm$^2$/n, respectively. The corresponding hard failure cross section around $2\times10^{-11}$ cm$^2$/n. During the whole evaluation, we did not observe any other abnormal states in current consumption for both setups.

## VI. DISCUSSION

*A. ATLAS NSW Router Data Loss Estimations*

The neutron beam energy spectra at LANSCE is quite similar to the neutron background present in the ATLAS Liquid Argon (LAr) frontend electronics [24]. Based on the relative locations between LAr and NSW detectors in the ATLAS experiment, it is expected that LAr experiences more radiation than NSW. Since there are no similar documents for the NSW detector at this moment, the radiation level at the LANSCE facility represents the worst case estimation for the NSW Router.

The Router is a crucial component in the trigger path in the NSW detector. Data loss is a direct concern in the radiation environment. Data loss is estimated by defining the data loss as a percentage (DL%) of abnormal interruption time ($T_{down}$) over 10 years of detector operation $(3.15 \times 10^8$ s$)$. In estimating $T_{down}$, we assume soft failures take 8 s to be fixed whereas hard failures require 60 s with an additional 300s required for each for failure confirmation. Thus:

$$T_{down} \approx [\sigma_{link\ (soft)} \times (8+300) + \sigma_{link\ (hard)} \times (60+300)] \times \Phi_{fluence} \quad (3)$$

We used the results in SET I, RUN 6, from the LANSCE experiment for the calculation, in which $\Phi_{fluence} \approx 1.04 \times 10^{12}$ n/cm$^2$ from Part II.A, $\sigma_{link\ (soft)} \approx 3.68 \times 10^{-10}$ cm$^2$, $\sigma_{link\ (hard)} \approx 2 \times 10^{-11}$ cm$^2$. With these, $T_{down} \approx 125366$ s. DL% is then estimated as: DL%=125366/315360000=0.04%, which is acceptable for the NSW trigger electronics.

### B. SEU Cross Section versus Beam Energy

SEE characterization tests are typically performed with mono-energetic proton cyclotron test facilities (e.g., PSI, protons up to 230 MeV) or low energy neutron facilities (e.g., NSCR, up to 20Mev) [25]. However, the radiation environment in the ATLAS experiment is complicated and includes radiation from a broad range of particles and energies. It is thus necessary to understand the failure probability for different particle types and energies. In our evaluation, we observed a significant contribution due to the neutron beam energy itself. This is reflected by the difference in the failure cross section between RUN 4, NSCR and RUN 5, LANSCE, in which the soft failure cross section of the latter is almost tripled. NSCR and LANSCE are both neutron facilities with similar fluxes in our tests. With identical setups and checking criteria, the only difference is the energy spectra. It is thus evident that the results from LANSCE evaluation should be more accurate that those from NSCR in estimating the final link loss percentage for the ATLAS NSW Router.

### C. Logic Utilization

A Xilinx Artix-7 FPGA (XC7A200TFFG1156-2) has been chosen for the Router implementation. This choice was driven by the relatively low cost of the Artix-7 FPGA family and by the total number of GTP transceivers available in this device. Table V summarizes the logic utilization of the FPGA. Less than 20% of the programmable logic resources (Slice Register and LUT) are used for the Router logic with TMR. In contrast, for dedicated resources such as the clocking buffer (BUFG), more than 50% is utilized. The Router implementation also uses 12 out of 16 GTPs, for the 12 incoming links and 4 outputs links. Each GTP consists of a RX and a TX for the incoming and outgoing links respectively.

### VII. CONCLUSION

The Router in the ATLAS NSW detector plays an important role in processing the giga-bit-per-second serial links in the trigger chain. It utilizes a commercial Xilinx Aritx-7 FPGA for serial link switching which is vulnerable to radiation induced effects in the NSW detector environment. We presented a multi-layer SEU mitigation scheme to protect the FPGA design from radiation effects, in which a combination of both circuit level (TMR and SEM IP) and system level (mBAR and power cycling) strategies were applied. We built test setups and demonstrated the FPGA performance at two different neutron facilities. Results show that the scheme is effective in mitigating SEUs and keeps the data loss of the Router design within acceptable levels. The implementation of the mitigation scheme in the Router represents a typical FPGA application and could be adapted to enhance the reliability of FPGA designs for similar applications in harsh radiation environments.

TABLE V
LOGIC UTILIZATION FOR ROUTER LOGIC

| Resource | USED/AVAILABLE | Utilization |
|---|---|---|
| Slice Register | 27437/267600 | 10.25% |
| Slice LUT | 24534/133800 | 18.34% |
| BUFG | 18/32 | 56.25 % |
| DCM/PLL | 4/10 | 40% |
| GTP | 12/16 | 75% |

### ACKNOWLEDGMENT

The authors would like to thank Michael Wirthlin from Brigham Young University (BYU), Helio Takai, James Kierstead from the Brookhaven National Lab (BNL), Stephen Wender from Los Alamos Neutron Science Center (LANSCE), and Edward Diehl from University of Michigan for their help in this work.